\documentclass[a4paper,11pt]{article}
\usepackage{jinstpub} 
\usepackage{lineno}

\title{\boldmath Picosecond resolution photoelectron emission lifetime detection system }

\author[a]{V.~Kakoyan,}
\author[a,1]{S.~Zhamkochyan,\note{Corresponding author.}}
\author[a]{S.~Abrahamyan,}
\author[a,g] {A.~Aprahamian,}
\author[a]{H.~Elbakyan,}
\author[a]{A.~Ghalumyan,}
\author[a]{A.~Kakoyan,}
\author[a]{S.~Mayilyan,}
\author[a]{A.~Papyan,}
\author[a]{H.~Rostomyan,}
\author[a]{A.~Safaryan,}
\author[a]{G.~Sughyan,}
\author[a]{N.~Margaryan,}
\author[b]{J.~Annand,}
\author[b]{K.~Livingston,}
\author[b]{R.~Montgomery,}
\author[c]{P.~Achenbach,}
\author[d]{J.~Pochodzalla,}
\author[e]{D.L.~Balabanski,}
\author[f]{S.N.~Nakamura,}
\author[g]{K.~Manukyan,}
\author[h]{V.~Sharyy,}
\author[h]{D.~Yvon,}
\author[a]{A.~Margaryan}
\affiliation[a]{A.I.~Alikhanyan National Science Laboratory (Yerevan Physics Institute),\\
2 Alikhanyan Brothers., Yerevan, Armenia}
\affiliation[b]{School of Physics \& Astronomy, University of Glasgow,\\
G12 8QQ Scotland, UK}
\affiliation[c]{Jefferson Laboratory,\\
Newport News, VA, USA}
\affiliation[d]{Institut für Kernphysik, Johannes Gutenberg-Universität Mainz,\\
Mainz, Germany}
\affiliation[e]{Extreme Light Infrastructure- Nuclear Physics (ELI-NP),\\
Bucharest-Magurele, Romania}
\affiliation[f]{Department of Physics, Graduate School of Science, the University of Tokyo,\\
Tokyo, Japan}
\affiliation[g]{Department of Physics and Astronomy, University of Notre Dame,\\
Notre Dame, IN 46556, USA}
\affiliation[h]{Département de Physique des Particules Centre de Saclay,\\
91191 Gif-sur-Yvette Cedex, France}

\emailAdd{szh@mail.yerphi.am}

\abstract{This paper describes a new photoelectron emission lifetime detection system. It is based on a recently developed Radio Frequency Timing technique of keV electrons and a 40 MHz, ultrafast pulsed laser. The photoelectron emission lifetimes from gold, monolayer MoS$_2$ and monolayer graphene were measured. As expected, we do not observe delayed electrons from gold, and the time distribution of the produced photoelectrons represents the time resolution of the device, which is $\sim$12~ps. From the graphene, we observed delayed photoelectrons with a lifetime of $\sim$189~ps.}


\begin{document}
\maketitle
\flushbottom

\section{Introduction}
\label{sec:intro}
The lifetime of hot carriers generated upon photoexcitation is critical in determining the feasibility of using materials in applications such as solar energy conversion and surface chemistry (see ~\cite{lin_fan_2018_gr1}). For instance, graphene, owing to its unique band structure, has emerged as a highly promising material for optoelectronic applications. Theoretically, it is expected to have long hot carrier lifetimes, ranging from hundreds of picoseconds to a few nanoseconds, due to its high optical phonon energy ($\sim$200~meV) and limited phase space for acoustic phonon scattering ~\cite{bistritzer_2009_gr2}. However, numerous experimental studies over the past decade using various pump-probe techniques (see ~\cite{lin_fan_2018_gr1} and references therein)  consistently report much shorter decay times of only a few picoseconds. This discrepancy is largely due to the use of high excitation energies in the near-IR and mid-IR regions, which result in efficient carrier-carrier scattering and rapid thermalization of electrons to temperatures of several thousand degrees Kelvin. At such high temperatures, optical phonon emission dominates the decay process. Furthermore, when electron temperatures fall below the optical phonon energy, a mechanism known as "supercollision"—involving multiple scatterings with defects that facilitate momentum transfer to acoustic phonons—has also been proposed to explain the fast decay rates~\cite{betz_2012_gr17,graham_2012_gr18}. In addition to graphene, other two-dimensional (2D) materials such as MoS$_{2}$ (molybdenum disulfide) have also shown promise due to their strong light-matter interaction, tunable band structure, and relevance to high-speed electronics, infrared photodetection, and even space research ~\cite{chengyun_2025_mos2_3}.

A breakthrough came with the development of a new Time-of-Flight Angle-Resolved Photoemission Spectroscopy (TOF-ARPES) system, which offers high detection efficiency and eliminates the need for conventional pump-probe setups ~\cite{lin_fan_2018_gr1}. When applied to copper films, this system measured a decay constant of 64~ps, which indicates the lower limit of decay time measurement of the system, since copper itself shows no delayed emission. Using this new technique  at very low laser fluence ($\sim$10~$\mu$J/cm$^2$), researchers were able to directly observe a much slower decay process in graphene at room temperature — about two orders of magnitude longer than previous measurements. This extended lifetime is believed to result from the excitation of image potential states, which are known to possess inherently long lifetimes. However, detecting such signals at low fluence is technically challenging due to the weak signal strength, rendering many conventional methods ineffective. These findings raise compelling questions about the potential role of image potential states at even lower fluences and whether they can be efficiently harvested for optoelectronic applications.

In this paper, we present a novel photoelectron emission lifetime detection system based on a recently developed high-time-resolution radio frequency (RF) timing technique for keV electrons ~\cite{rftimer_nim, rftimer_jinst}. This approach enables lifetime measurements using extremely low laser fluences — on the order of a few nJ/cm$^2$. 

\section{Photoelectron emission lifetime spectrometer}
\label{sec:spectrometer}
\begin{figure}[htbp]
\centering
\includegraphics[width=0.85\textwidth]{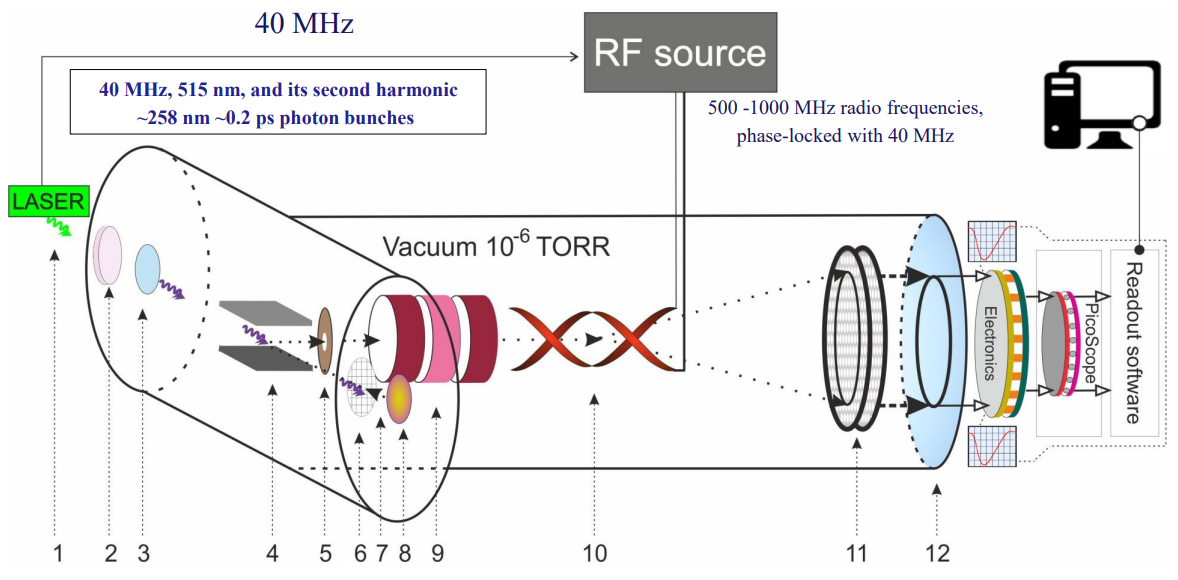}
\caption{Schematic of the spectrometer. 1-laser photon pulse; 2-nonlinear crystal; 3-quartz window; 4-magnet; 5-collimator; 6-accelerating electrode; 7-photoelectron; 8-cathode or material under study; 9-electrostatic lens; 10-RF deflector; 11-MCP detector; 12- delay line anode.\label{fig:setup_spectrometer}}
\end{figure}

The photoelectron emission lifetime spectrometer is schematically shown in figure~\ref{fig:setup_spectrometer}. It consists of several parts: an ultrafast 40 MHz laser, RF timing tube, RF synthesizer, position-sensing detector and electronics, data acquisition system, power supplies, and vacuum system. The RF timer ~\cite{rftimer_nim, rftimer_jinst} components are mounted in a tube and operated at a vacuum of $10^{-6}$~Torr.

\begin{figure}[htbp]
\centering
\hspace{-3mm}
\includegraphics[width=.29\textwidth]{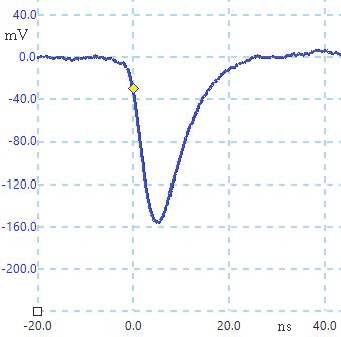}
\includegraphics[width=.29\textwidth]{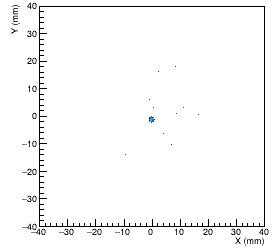}
\includegraphics[width=.29\textwidth]{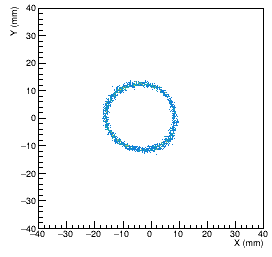}
\caption{(Left) typical amplified signal from the DL anode; (middle) 2D image of the focused electrons (RF is OFF); (right) 2D image of the scanned electrons (RF is ON).\label{fig:signal}}
\end{figure}

To test the system, we have used a thermionic emitter of electrons, specifically the model ES-042 Tantalum Disc Cathode from Kimball Physics. The diameter is 0.84~mm, and the work function is 4.1~eV.  The cathode, heated to 2200~K, emits a continuous low-energy electron beam with an energy spread of $\sim$0.6~eV. The emitted electrons are accelerated by a voltage of $\sim$2.5~kV applied between the cathode (8) and an accelerating electrode (6), which is a copper disc with a 1.5 mm diameter hole in the center, located $\sim$3~mm from the cathode. The accelerated electrons are deflected through 90 deg. by the permanent magnet (4) and pass through a collimator (5) with a diameter of $\sim$1.0~mm, before entering the electrostatic lens (9). This focuses the electrons onto the position sensitive detector (PSD), which consists of a dual chevron microchannel plate (MCP) (11) and a delay line (DL) anode (12). However, before reaching the PSD, the electrons pass through the RF deflection system (10) consisting of the half period, helical electrodes~\cite{rftimer_theory}, and a 520 MHz RF synthesizer. The deflector has a diameter of 8~mm and a length of 30~mm, and its exit point sits $\sim$120~mm from the MCP system. The electrons are multiplied by a factor of $\sim10^6$ in the MCP system, and the resulting electron cloud hits the DL anode, producing position-sensitive signals with rise times of a few ns. The DLD40 DL anode ~\cite{delay_line}  is used, which is a high resolution 2D imaging and timing device with 25~mm or 40~mm diameter active MCP area. Hit coordinates are sensed by differential wire pairs, formed by a collection (signal) wire and a reference wire. The four wire pairs feed through the vacuum container for external processing. These signals are then amplified by custom-built fast amplifiers and sent to the Data Acquisition (DAQ) system, which consists of digitizing oscilloscope (such as PICOSCOPE), running software to reconstruct the electron hit position on the MCP. The position of the detected electrons is encoded in the difference signal arrival times for each end of each parallel-pair delay line. A typical amplified signal from the DL anode, along with 2D images of the focused electron beam for RF power OFF and RF power ON, are shown in figure~\ref{fig:signal}.

The pickup on the anode induced by the RF power is negligible and does not affect the reconstruction of the image (time) of the RF-scanned electrons. Subsequent test studies were carried out with UV photons. The work function of the Tantalum cathode is 4.1~eV, so that photons with energies exceeding 4.1~eV (wavelength < 302~nm) can produce electrons. The vacuum tube of the spectrometer has a quartz-glass window (2 in figure~\ref{fig:setup_spectrometer}), which also allows testing of this new timing technique using UV photons. The studies started with a UV diode, operating at 273~nm (4.54~eV) in CW mode. Photons from the diode transmit through the quartz glass window, hit the cathode, and produce photoelectrons (PE) which are transported through the RF timer as in the test with thermo-electrons. Typical signals from the anode and the obtained hit-position distributions, in cases where the RF source is OFF and ON, are very similar to those obtained with thermoelectrons. In the next step, experimental studies continued using the RF-synchronized laser.

\section{RF-synchronized laser}
\label{sec:laser}
The optical part of the spectrometer consists of an ultra-low noise femtosecond laser module ORIGAMI-05LP from NKT. The laser module provides 166~fs duration pulses at a wavelength of 515~nm, with a repetition rate of 40~MHz and an average power of 25~mW. A  Beta Barium Borate (BBO) non-linear crystal shifts the 515~nm wavelength of the laser beam by a factor of 2 down to 257.5~nm. After the BBO, the laser beam was filtered using a UF-5 optical filter, and 257.5~nm photon bunches with 166~fs duration and 40~MHz repetition rate were directed on the material under study. The fluence of the UV photon beam on the material under study was ~2.5 nJ/cm$^2$. The laser module also provides electrical signals of ns duration synchronized with the photon bunches. These signals, after appropriate filtering, were used by a radio frequency synthesizer to generate a 500-1000~MHz RF power source, phase-locked to 40~MHz, which was used to power the radio frequency deflector. For the present studies the RF was set to 520~MHz.

\section{Operation of the photoelectron emission lifetime spectrometer with the RF-synchronized laser}
\label{sec:operation}
Photon pulses at 515 and 257.5~nm, generated as described in Sec.3, were directed at the different materials under study. The RF signal $V_{\mathrm{RF}}(t)$ can be written in the form:
\begin{equation}
    \label{eqn:1}
    V_{\mathrm{RF}}\left(t\right)=V_{\mathrm{RF}}^0\mathrm{sin}[2\pi\nu_{\mathrm{RF}}t + \varphi_{\mathrm{RF}}\left(t\right) + \varphi_{\mathrm{RF}}^{0}],    
\end{equation}
where the amplitude $V_{\mathrm{RF}}^0$ is assumed constant. The quantity $\nu_{\mathrm{RF}}=n\times\nu^0$, where $\nu^{0}$ is the constant nominal frequency of the mode-locked laser, n is an integer and t is the time referenced to the laser pulse. The phase $\varphi_{\mathrm{RF}}\left(t\right)$ contains the random and systematic deviations, relative to the ideal oscillation $\nu^0$. In our case $\nu^0=\mathrm{40 MHz}$ and $n=\mathrm{13}$ so that $\nu_{\mathrm{RF}}=\mathrm{520} MHz$. The quantity $\varphi_{\mathrm{RF}}^0$ represents the nominal phase, which is constant for a given setup.
It is expected that the RF signal in the equation \ref{eqn:1} has the same time structure as the laser photon beam with the same random and systematic deviations, $\varphi_{\mathrm{RF}}\left(t\right)$. Photoelectrons produced in the material under study are accelerated, deflected by the magnet, and focused on the PSD by the electrostatic lens.  They pass through the RF deflector at a time  $t^i$ and fix the total phase of the RF oscillator at a point  on the scanning circle, for a given $\varphi_{\mathrm{RF}}^0$:
\begin{equation}
    \label{eqn:2}    \Phi_{\mathrm{RF}}^i=2\pi\nu_{\mathrm{RF}}^{0}t^i+\varphi_T\left(t^i\right)+\varphi_{\mathrm{RF}}^0,    
\end{equation}
Here the phase $\varphi_T\left(t^i\right)$ contains random and systematic deviations relative to the ideal oscillation, due to the spectrometer. The time $t^{i+1}$ for the next photon will be transposed to the phase:
\begin{equation}
    \label{eqn:3}    \Phi_{\mathrm{RF}}^{i+1}=2\pi\nu_{\mathrm{RF}}^{0}t^{i+1}+\varphi_T\left(t^{i+1}\right)+\varphi_{\mathrm{RF}}^0    
\end{equation}
For an ideal device, without any random or systematic drifts $\left(\varphi_T\left(t\right)=0\right)$, the produced PEs are perfectly in phase with the photon bunches. In this case, the sinusoidal RF signal from the laser plays the role of time reference. The timing jitter of the spectrometer may be measured from the PEs produced by the incident photon pulses   on a metal, e.g. gold, for which photoelectron emission is prompt. The radius R and phase $\varphi$ of the scanned electrons were determined by transforming the measured X,Y from the PSD to polar coordinates and the resulting $\varphi$ was then converted to ns~\cite{rftimer_nim} on the basis that 360 deg. is equivalent to 1000/520~ns.

\begin{figure}[htbp]
\centering
\includegraphics[width=0.85\textwidth]{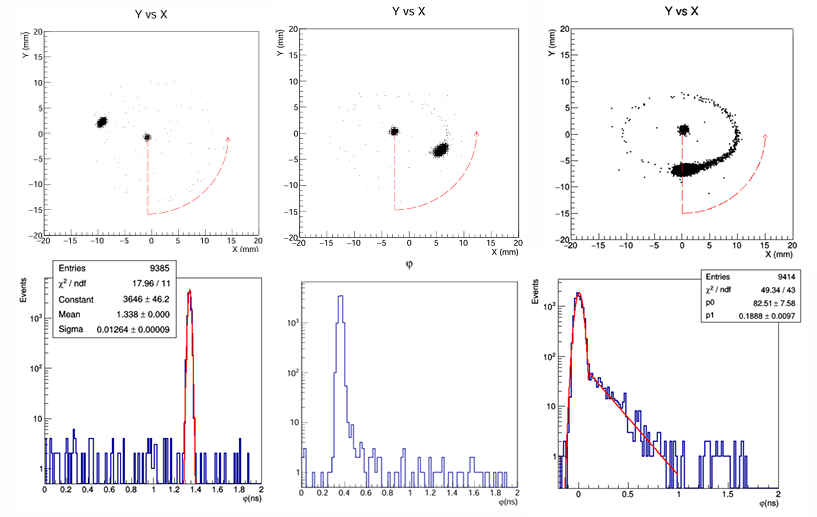}
\caption{Results obtained from gold (left),
monolayer MoS$_2$ (middle), and monolayer graphene (right): (top) 2D images of anode hit positions, (bottom) phase distributions of the 520 MHz scanned synchronized photoelectrons. \label{fig:results}}
\end{figure}

Figure~\ref{fig:results} shows the 2D image of anode hit positions (top), as well as $\varphi$ distributions (bottom), for a fixed phase of RF synchronization to the laser photon beam, for gold (left), monolayer MoS$_2$ (middle), and monolayer graphene (right). The spot in the center of figures is a 2D image of the 2.5~keV electrons, obtained when the RF is OFF.

The 2D image of scanned electrons corresponds to the phase distribution of photoelectrons produced by RF-synchronized, 257.5~nm laser photons, the fluence of which is about 2.5~nJ/$\mathrm{cm}^2$. In the case of gold, all photoelectrons have the same phase, and a spot on the circle is obtained. The spread in phase of the spot represents the overall time resolution of the system, which includes factors related to the laser, the laser and RF oscillator synthesizer device, and the intrinsic time resolution of the spectrometer. It follows from the $\varphi$ distribution of photoelectrons from gold, presented in figure~\ref{fig:results}, that the time resolution (sigma of the Gaussian fit) is 12.6$\pm$0.1~ps. In the case of MoS$_2$, we recorded delayed photoelectrons, which produced a slight tail to the right of the prompt peak. In the case of graphene, there is a more substantial tail. We fit the prompt part of the graphene spectrum with the Gaussian function and then fit the tail (starting 3 sigma from the Gaussian mean) with an exponential function. The result of the fit corresponds to a lifetime of 189$\pm$10~ps.

These results demonstrate that the system is capable of performing precise photoemission time measurements using a low-intensity excitation light source, opening new opportunities for studying hot carriers in two-dimensional materials.

\section{Summary and outlook}
\label{sec:summary}
This paper describes a new photoelectron emission lifetime detection system. It is based on a recently developed Radio Frequency Timing technique of keV electrons and a 40~MHz, ultrafast pulsed laser. The photoelectron emission lifetimes from gold, MoS$_2$, and graphene were measured. As expected, we do not observe delayed electrons from gold, and the time distribution of the produced photoelectrons presents the time resolution of the device, which is 12.6$\pm$0.1~ps. From the graphene, we observed delayed photoelectrons with a lifetime of 189$\pm$10~ps. Experimental studies with this ultrahigh precision technique are continuing. The 12.6~ps resolution is mainly due to the technical parameters of the device. With a view to achieving ps resolution or better, we will continue to optimize these parameters, e.g. by improving synchronization between the laser and RF, and using dedicated high-frequency deflectors. Eventually, a few hundred fs time resolution is expected for single electrons with such an optimized device, operated with 10~GHz RF.

\acknowledgments
This work was partially supported by the Higher Education and Science Committee of the Republic of Armenia (Research Project: 23LCG-1C018), the International Science and Technology Center (ISTC project AM-2803), the ARPA Institute and the UK Science and Technology Facilities Council. The work of P.A. is supported by the U.S. Department of Energy, Office of Science, Office of Nuclear Physics under contract DE-AC05-06OR23177. The work of D.L.B. is partially supported by the Romanian
Ministry of Research, Innovation and Digitalization under Contract No.
PN 23 21 01 06 an ELI-RO project Contract ELI-RORDI-2024-007 (ELITE).

\bibliographystyle{JHEP}
\bibliography{biblio.bib}

\end{document}